\documentclass[%
 reprint,
 amsmath,amssymb,
 aps,
]{revtex4-2}

\usepackage{graphicx}
\usepackage{dcolumn}
\usepackage{bm}
\usepackage{mathrsfs}
\def \Dperp {\overset{\perp}{\nabla}}
\usepackage{amssymb, amsmath}

\usepackage{tikz-cd}
\usepackage{comment}
\usepackage{balance}

\begin{document}

\title{Non-inertial frames that can \textit{mimic} gravitational fields}

\author{Dawood Kothawala}
 \email{dawood@iitm.ac.in}
\affiliation{Indian Institute of Technology Madras, Chennai}%

\date{\today}

\begin{abstract}

\noindent One version of the principle of equivalence, as originally formulated by Einstein, states that ``gravity" can be mimicked locally by going to an ``accelerated frame of reference". As highlighted by Synge, the physical content of this principle remains obscure in so far as it does not refer to the Riemann tensor $R_{abcd}$, which encodes the true effects of gravity.
%
%
We here give the acceleration profile of a \textit{Born} rigid, Rindler{\it esque}, frame that can mimic a gravitational field corresponding to a given $R_{abcd}$. 
The generalised deviation equation that yields this result also has Centrifugal and Coriolis terms appearing in a purely relational context, yielding a similar connection between angular velocity of rotating, rigid inertial frames and Riemann. We comment briefly on implications for Mach principle.
\end{abstract}

\maketitle


\noindent \textit{\textbf{Introduction:}} Almost every introduction to general relativity begins with the statement of the so called \textit{principle of equivalence}, which, in its simplest form, states that one can deduce physical effects in presence of gravity by instead working in an accelerated frame of reference in flat spacetime without gravity. Delving deeper, one then encounters this principle at various increasing levels of sophistication. Our focus in this work, however, will be the version stated above. Although simple to state and easy to visualize, when probed deeper, one encounters subtleties concerning the actual physical content of this statement EP. These were most prominently highlighted by Synge, in the preface of his classic text on general relativity \cite{synge-gr}. The key point is this: if accelerated frames in Minkowski spacetime are supposed to mimic effects due to gravity, and if true gravitational effects are encoded in the Riemann tensor $R_{abcd}$, the EP is void unless one also shows how the ``acceleration" of this frame determined by ``curvature". For instance, one can compute the acceleration of objects near the surface of the Earth to be $9.8\, {\rm m/s^2}$, and hence, following EP, do physics near the Earth's surface by going to an accelerated frame with the above reference value of acceleration. How is this value derivable from the curvature produced by the Earth? Our main aim in this work is to present a derivation which clarifies these issues in a quantitative manner, thereby giving a precise meaning to the physical content of the EP.
\\\\
\noindent \textit{\textbf{Rigid congruences in general relativity:}} Consider a congruence of timelike curves $C_n$ in an arbitrary curved spacetime, parameterised by a real parameter $n$, with four velocity $u^a$ and proper time $\tau$ (both parameterised by $n$). We will assume $u^a u_a=-1$. (Latin indices are spacetime indices, whereas Greek indices represent space indices.) Let $C_0$ be one of these family of curves, which we will call as the reference curve. Let $\Sigma$ be a spacelike hypersurface on which we set $\tau=0$ for all congruences. One may then define a connecting vector $\xi^a$, which connects events on two members of the congruence, at the same proper interval from $\Sigma$. Being a connector, such a vector field will satisfy $\mathscr{L}_{\bm u} {\bm \xi}=0$, where $\mathscr{L}_{\bm u}$ represents the Lie derivative along the vector field ${\bm u}$ (all bold face symbols such as $\bm u, \bm{x}, \bm{v}, \bm{g}, \bm{Z},$ etc represent spacetime vector fields). A good measure of deviation between members of the congruence is then obtained from the projection of the above vector orthogonal to $u^a$: $Z^a = \xi^a + u^a (\xi^k u_k)$. Let every member of the congruence be equipped with a tetrad that transports itself according to the generalised Fermi-Walker transport described by the derivative operator \cite{mtw}
\begin{eqnarray}
    \frac{\widetilde D_F q^i}{d \tau} = \nabla_{\bm u} q^i + (a^i u_j - u^i a_j) q^j - {\mathbb R}^i_{\phantom{i} j} q^j
\end{eqnarray}
where $\nabla_{\bm u} \equiv u^k \nabla_k$, $a^i = \nabla_{\bm u} u^i$ is the acceleration, and ${\mathbb R}^i_{\phantom{i} j}$ is a rotation matrix characterising arbitrary rotation of the spatial tetrads; ${\mathbb R}^i_{\phantom{i} j} u^j=0={\mathbb R}^i_{\phantom{i} j} u_i$. A (generalized) Fermi tetrad basis ${\bm e}_{(i)}$ is then defined by the transport law
\begin{eqnarray}
    \frac{\widetilde D_F {\bm e}_{(i)}}{d \tau} = 0
\end{eqnarray}
with ${\bm e}_{(0)}=\bm u$ \cite{mtw}. 

With these definitions, once can attach notions such as relative velocity and relative acceleration between members of the congruence. These are given, simply, by 
\begin{eqnarray}
    \bm v &=& \frac{\widetilde D_F \bm Z}{d \tau}
     \\
    \bm g &=& \frac{\widetilde D_F \bm v}{d \tau}
\end{eqnarray}
Explicit expressions for these can be obtained through a straightforward, but somewhat long, calculations. We have sketched the steps in the supplemental material. The calculations essentially follow a route similar to the conventional derivation of the equation of \textit{geodesic deviation}, but tailored for congruences whose members can have arbitrary acceleration. The final form for these (see Supplemental Material) turn out to be
\begin{widetext}
    \begin{eqnarray}
    v^a &=& \left( K^a_{\phantom{a}b} - {\mathbb R}^a_{\phantom{a}b} \right) Z^b
    \label{eq:v}
    \\
    g^b &=& \Dperp_{\bm u} \Dperp_{\bm u} Z^b + 
    \left( \left({\mathbb R}^2\right)^a_{\phantom{a}m} - 2 {\mathbb R}^a_{\phantom{a}b} K^b_{\phantom{b}m} \right) Z^m - e^a_{(\alpha)} Z_\beta \dot{\mathbb R}^{\alpha \beta}
    \label{eq:g1}
    \end{eqnarray}
\end{widetext}
with the definitions $K^a_{\phantom{a}b} = \nabla_b u^a + u_b a^a$, $\left({\mathbb R}^2\right)^a_{\phantom{a}m} = {\mathbb R}^a_{\phantom{a}b} {\mathbb R}^b_{\phantom{b} m}$, and $Z_\beta$ and ${\mathbb R}^{\alpha \beta}$ are spacetime scalars representing tetrad components of $Z^a$ and ${\mathbb R}^a_{\phantom{a}b}$, with $\dot{\mathbb R}^{\alpha \beta} = d{\mathbb R}^{\alpha \beta}/d\tau$. $\Dperp$ represents covariant derivative projected perpendicular to $u^a$. The first term on the RHS of Eq.~(\ref{eq:g1}) is the term familiar from conventional derivation of deviation equations, and is given by 
    \begin{eqnarray}
    \Dperp_{\bm u} \Dperp_{\bm u} Z^b &=&  
    - {\mathscr E}^b_{\phantom{b}m} Z^m + \left( Z^m a_m \right) a^b + 
    \left( \nabla_{\bm Z} a^m \right)_\perp
    \end{eqnarray}
where ${\mathscr E}^a_{\phantom{a}b} = R^a_{\phantom{a}ibj} u^i u^j$ is the electric part of Riemann, and the last term on RHS stands for the projection of the vector $\nabla_{\bm Z} a^m$ orthogonal to $u^a$. 

Using the above, and Eq.~(\ref{eq:v}), one may write Eq.~(\ref{eq:g1}) in a very suggestive form (we have restored the speed of light $c$ for purpose of comparison with Newtonian terms):
\begin{widetext}
    \begin{equation}
    g^b =
    - c^2 {\mathscr E}^b_{\phantom{b}m} Z^m + 
    \underbrace{
    \vphantom{\left( \frac{Z^m a_m}{c^2} \right)}
    \left( \nabla_{\bm Z} a^b \right)_\perp}_{\rm gradient} + 
    \underbrace{\left( \frac{Z^m a_m}{c^2} \right) a^b}_{\rm relativistic ~ redshift} 
    - \; \underbrace{
    \vphantom{\left( \frac{Z^m a_m}{c^2} \right)}
    \left({\mathbb R}^2\right)^b_{\phantom{b}m} Z^m}_{\rm Centrifugal} 
    \; - \; 
    \underbrace{
    \vphantom{\left( \frac{Z^m a_m}{c^2} \right)}
    2 {\mathbb R}^b_{\phantom{b}m} v^m }_{\rm Coriolis} 
    \; - \; 
    e^b_{(\alpha)} Z_\beta \dot{\mathbb R}^{\alpha \beta} 
    \label{eq:g2}
    \end{equation}
\end{widetext}
The above expression reduces to known ones in appropriate limits. For geodesic congruences and non-rotating tetrads, $a^b=0={\mathbb R}^a_{\phantom{a}b}$, only the first term on the RHS survives and we recover the standard textbook version of the equation of geodesic deviation. For accelerated congruences with non-rotating tetrads, ${\mathbb R}^a_{\phantom{a}b}=0$, and we recover Eq. (4.4) of Hawking and Ellis \cite{hawking-ellis}. Equation (\ref{eq:g2}) will be the key equation on which rest of the analysis in this work will be based. 
In fact, it will be enough to work with the scalar form of this equation that governs time evolution of the magnitude of $Z^b$. For this purpose, we write $Z^b=Z n^b$, where $n^b$ is a spacelike vector field of unit norm ($n^a n_a=1$) orthogonal to $u^a$, so that $u^a n_a=0$, and $Z$ is the magnitude of $Z^b$: $Z^b Z_b=Z^2$. Following the procedure sketched in Appendix A of \cite{dk-bgv}, we then obtain:
\\ \\
\begin{widetext}
    \begin{equation}
    \frac{1}{Z} \frac{d^2 Z}{d \tau^2} =
    - c^2 {\mathscr E}_n + \frac{\partial a_n}{\partial n} + \frac{a_n^2}{c^2} - \bm a \cdot \nabla_{\bm n} \bm n
    + \left[ \left({\mathbb R}^2\right)^a_{\phantom{a}m} - 2 {\mathbb R}^a_{\phantom{a}b} K^b_{\phantom{b}m} \right] n_a n^m
    + \left(\frac{\widetilde D_F \bm n}{d \tau} \right)^2
    \label{eq:gmag}
    \end{equation}
\end{widetext}
where ${\mathscr E}_n = {\mathscr E}^a_{\phantom{a}b} n_a n^b$ and $a_n=\bm a \cdot \bm n$. As we will see, the above equation is key to understanding how spacetime curvature can inform a choice of 
reference frame, and how these relate to the \textit{equivalence principle}.
\\\\
\noindent \textit{\textbf{Accelerated frames and curvature:}} Equation (\ref{eq:gmag}) clearly highlights the different contributions to what might be the most convenient \textit{operational} way of quantifying relative acceleration between curves in general. Terms involving $\{a^i, {\mathbb R}^a_{\phantom{a}b} \}$ are basically related to the observer frame tied to each member of the congruence $C_n$, while the $\{ \partial_n a_n, K^a_{\phantom{a}b}\}$ terms depend on the entire congruence. Note that $K^a_{\phantom{a}b}$ appears only through its coupling to ${\mathbb R}^a_{\phantom{a}b}$, and does not appear for a tetrad which is \textit{Fermi} transported. Finally, there is the universal contribution from spacetime curvature through the tidal part of Riemann, ${\mathscr E}_n$.

Consider, then, a non-rotating \textit{frame} (${\mathbb R}^a_{\phantom{a}b}=0$) which is constructed so that $\ddot Z=0$. This means that the members of the congruence that are used as frame axes do not accelerate with respect to each other, and no stresses are required to maintain their relative orientation
\footnote{
The consequences of such a condition for rigidity have been discussed in the literature previously (see, for example, \cite{synge-gr} and \cite{pirani-rigidity}), although these discussions mostly work only with $\dot Z$ and not its second derivative. 
}.
We will also assume that the orthogonal separation is along geodesics, so that $\nabla_{\bm n} \bm n=0$. This seems like a reasonable choice, but we will comment on it further in the last section. What is the acceleration profile that corresponds to this frame? Choose a direction which is \textit{fixed} in this frame, so that ${\widetilde D_F \bm n}/{d \tau}=0$. From Eq. (\ref{eq:gmag}), the acceleration profile along $\bm n$ can then be obtained from the equation
\begin{eqnarray}
    \frac{\partial a_n}{\partial n} + \frac{a_n^2}{c^2} = c^2 {\mathscr E}_n
\end{eqnarray}
This is a Riccati equation for $a_n$, and one can obtain an exact solution of it for a constant ${\mathscr E}_n$. That is, if one ignores derivatives of Riemann, $\nabla R \sim 0$, the solution is 
\\
\\
\begin{equation}
    a_n =  c^2 \sqrt{{\mathscr E}_n} \tanh \left[ \sqrt{{\mathscr E}_n} n +  \tanh^{-1}{\left(\frac{a_0}{\sqrt{{\mathscr E}_n}}\right)} \right]
    \label{eq:an-sol}
\end{equation}
\\
\\
where $a_0$ is the (normal component of) acceleration for $C_0$ - the reference curve. 

\noindent \textit{\underline{Features of Rindler{\it esque} profiles in curved spacetime:}} From the above solution, one can deduce the following:
\begin{enumerate}
    \item In flat spacetime, we obtain 
    \begin{eqnarray}
        \lim \limits_{{\mathscr E}_n \to 0} a_n = \frac{a_0}{1+a_0 n}
        \label{eq:rindler}
    \end{eqnarray}
    For $a_0=0$, this is just the Minkowski inertial congruence, while for $a_0 \neq 0$, we recover the Rindler congruence. 

    \item In curved spacetime, for $a_0=0$, 
    \begin{eqnarray}
        \lim \limits_{a_0 \to 0} a_n = c^2 \sqrt{{\mathscr E}_n} \tanh \left[ \sqrt{{\mathscr E}_n} n \right]
    \end{eqnarray}
    which has the following ``asymptotic" behaviour
    \begin{eqnarray}
    \lim \limits_{a_0 \to 0} a_n =
    \begin{cases}
    \;\; c^2 {\mathscr E}_n n \;\;\;\; \left(\mathrm{for} \;\; n \ll 1/\sqrt{{\mathscr E}_n} \right)
    \\
    \\
    \;\; c^2 \sqrt{{\mathscr E}_n} \;\;\;\; \left(\mathrm{for} \;\; n \gg 1/\sqrt{{\mathscr E}_n} \right)
    \end{cases}
    \label{eq:cases}
    \end{eqnarray}
The physical meaning of these limits is the following: If one chooses the reference curve as a geodesic, then the \textit{acceleration} one needs to impart to the neighbouring curves to form a rigid frame is precisely equal to the geodesic deviation acceleration between the reference curve and a neighbouring \textit{geodesic}; this acceleration has a typical profile which is linear in $n$. Note, however, that this is true only as long as these curves lie within the local radius of curvature. On the other hand, beyond the curvature length scale, the acceleration needed asymptotes to a constant value fixed by curvature. 

    \item For $a_0 \neq 0$ and $n \ll 1/\sqrt{{\mathscr E}_n}$, 
    \begin{eqnarray}
        a_n = a_0 - \left(a_0^2 - {\mathscr E}_n \right) n + O(n^2)
    \end{eqnarray}
    The above expansion can be compared with that of Rindler profile in Minkowski, which can be obtained from Eq. (\ref{eq:rindler}) as: 
    $
        \widetilde{a}_n = \widetilde{a}_0 - \widetilde{a}_0^2 n + O(n^2),
    $
    where we have used tilde to highlight that these values are in Minkowski. The coefficient of $n$ in flat and curved spacetime has special significance and has been encountered in the context of Unruh effect in curved spacetime in \cite{hari-dk-acc-detectors}.
\\
\end{enumerate}

\begin{figure}[htb!]%
    {{\includegraphics[width=0.45\textwidth]{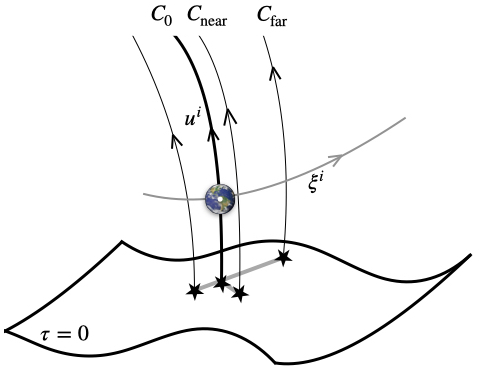} }}%
    \caption{The origin of \textit{accelerated frames} that can ``mimic" gravity in the vicinity of Earth (curve $C_{\rm near}$), in terms of Riemann tensor at the center of the Earth, on $C_0$, assumed to be on geodesic.}%
    \label{fig:rigid-frames}%
\end{figure}

\noindent \textit{\underline{Maximally symmetric spacetimes:}}
Our solution Eq.~(\ref{eq:an-sol}) for the acceleration profile is \textit{exact} for maximally symmetric spacetimes, for which $R_{abcd}=\Lambda \left(g_{ac} g_{bd} - g_{ad}g_{bc} \right)$, with $\Lambda=$ constant. This yields $\mathscr{E}_n = -\Lambda$. Therefore, we have $\mathscr{E}_n<0$ for de Sitter and $\mathscr{E}_n>0$ for anti-de Sitter spacetimes. We will highlight some salient features of the congruence for these spacetimes separately.
\\ \\
\noindent \textit{{(i) Anti-de Sitter:}}
For $\Lambda<0$, the entire discussion the previous paragraphs go through, except that the results are now exact. 
\begin{equation}
    a_n =  c^2 \sqrt{|\Lambda|} \tanh \left[ \sqrt{|\Lambda|} n +  \tanh^{-1}{\left(\frac{a_0}{\sqrt{|\Lambda|}}\right)} \right]
\end{equation}
from which it immediately follows that $|a_n|$ has an upper bound given by $c^2 \sqrt{|\Lambda|}$. Apart from this, there are no other constraints on the acceleration, and its different limits are as given in the previous paragraphs. 
\\ \\
\noindent \textit{{(ii) de Sitter:}} This case is more interesting case since it leads to certain new features. For $\Lambda>0$, the solution Eq.~(\ref{eq:an-sol}) leads to the following (exact) result
\begin{equation}
    a_n = - c^2 \sqrt{\Lambda} \tan \left[ \sqrt{\Lambda} \, n -  \tan^{-1}{\left(\frac{a_0}{\sqrt{\Lambda}}\right)} \right]
\end{equation}
which has several salient features. First, note that $-\infty < a_n < \infty$, and the values at which $a_n$ becomes $\pm \infty$ are given by 
    \begin{eqnarray}
    n_{\pm} = \mp \frac{\pi}{2\sqrt{\Lambda}} + \frac{1}{\sqrt{\Lambda}} \tan^{-1}{\left(\frac{a_0}{\sqrt{\Lambda}}\right)} 
    \end{eqnarray}
In the limit $\Lambda \to 0$ (using $\tan^{-1}x \sim (\pi/2) - x^{-1}$ for large $x$), these roots reduce to 
    \begin{eqnarray}
    n_{+} &\to& - \frac{1}{a_0}
    \nonumber \\
    n_{-} &\to& + \frac{\pi}{\sqrt{\Lambda}} 
    \end{eqnarray}
which are, respectively, the Rindler and the cosmological horizons. It is quite intriguing that these are derived here simply from the structure of the acceleration profile obtained from the rigidity condition alone.
\\\\
\noindent \textit{\textbf{The Equivalence principle:}} We are now ready to discuss the implications of the above for the equivalence principle. To take a concrete example, consider Fig.(\ref{fig:rigid-frames}). We assume the center to be moving on a geodesic, so that $a_0=0$, and also assume the size of the Earth (the reference frame) to be much smaller than the curvature scale at the center. The relevant limit is then the first of Eqs.~(\ref{eq:cases}), with $n=R_\oplus$, the radius of the Earth. In general, the matrix ${\mathscr E}^a_{\phantom{a}b}$ will be a combination of the contribution due to Earth, as well as from rest of the universe; to lowest order 
$$
{\mathscr E}^a_{\phantom{a}b} \sim {\mathscr E}^a_{\oplus b} + {\mathscr E}^a_{{\rm universe}\, b} 
\hspace{1cm} \mathrm{(at~ Earth's~ center)}
$$ 
For now, let us ignore the (potentially interesting!) contribution of ``rest of the universe" and focus on the contribution of Earth alone. Either by considering the Newtonian limit (valid since $GM_\oplus/c^2 R_\oplus \ll 1$), or assuming a Buchdahl constant density solution for the interior, one can easily show that the matrix ${\mathscr E}^a_{\phantom{a}b}$ has only one eigen-direction which satisfies $\nabla_{\bm n} \bm n=0$, which is $\bm n=\widehat{\bm r}$, the unit vector in the radial direction. At $r=0$, the corresponding eigenvalue is given by ${\mathscr E}_{\hat{r}}(0) = R_{\hat{0} \hat{r}\hat{0}\hat{r}}(0) = {G M_\oplus}/{R_\oplus^3}$. Note that 
$$\lim \limits_{r \to R_\oplus^+} {\mathscr E}_{\hat{r}} = {-2G M_\oplus}/{R_\oplus^3} \neq \lim \limits_{r \to R_\oplus^-} {\mathscr E}_{\hat{r}},$$ 
and in general 
${\mathscr E}_{\hat{r}}$ has a discontinuity at $r=R_\oplus$, both in sign and in magnitude, and hence our result can not be obtained merely from dimensional analysis. From the first of Eqs.~(\ref{eq:cases}), we then obtain, with $n \sim R_\oplus$,
\begin{eqnarray}
    a_{\hat r}(R_\oplus) &\sim& c^2 {\mathscr E}_{\hat{r}}(0) R_\oplus 
    \nonumber \\
    &=& {G M_\oplus}/{R_\oplus^2}
\end{eqnarray}
Outside (but near) to the Earth's surface, one can therefore:
\\
\\
\noindent ``\textit{mimic Earth's gravity by introducing a Rindler like frame with acceleration obtained directly from the Riemann tensor at the center.}" 
\\    

\begin{figure}[htb!]%
    {{\includegraphics[width=0.45\textwidth]{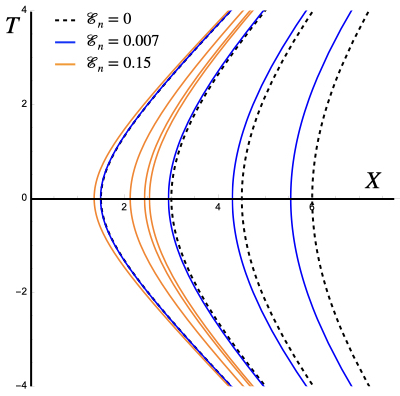} }}%
    \caption{The Rindler{\it esque} mimickers of curved spacetime. The acceleration of $C_0$ (not shown) is $a_0=2$. The difference from Rindler profile, both at high curvature and large distances, must be noted.}%
    \label{fig:rindler-curved}%
\end{figure}

To compare the obtained acceleration profile with the Rindler profile in Minkowski spacetime, we simply substitute $a_n$ in the relation $T=(1/a_n) \sinh[a_0 t], X=T=(1/a_n) \cosh[a_0 t]$, where $t$ is the proper time on $C_0$. The resultant congruence, depicted in Fig. (\ref{fig:rindler-curved}), highlights the subtleties, discussed above, that arise when using the equivalence principle in curved backgrounds. 
\\\\
\noindent \textit{\textbf{Rotation:}} We now look at Eq.~(\ref{eq:g2}) with rotation included, in which case it is interesting to note the appearance of \textit{fictitious-like} forces in a purely relational context. Note that the Coriolis term would disappear if one chooses ${\mathbb R}^a_{\phantom{a}b}$ such that $v^b=0$; this is only possible if the congruence is shear-free and one sets ${\mathbb R}^a_{\phantom{a}b}= K^a_{\phantom{a}b}$. This has intriguing consequence. For a shear free, geodesic congruence, the evolution equation for vorticity $\omega_{ab}$ (the anti-symmetric part of $K_{ab}$) would imply $\nabla_{\bm u} \omega_{ab}=0$ \cite{poisson-toolkit}, which implies
\begin{eqnarray}
    u^a \partial_a \omega_{\alpha \beta} = \omega_{\rho \beta} {\mathbb R}^\rho_{~\alpha} - \omega_{\sigma \alpha} {\mathbb R}^\sigma_{~\beta}
\end{eqnarray}
for the tetrad components. The RHS vanishes for ${\mathbb R}^\sigma_{~\beta} \propto \omega^\sigma_{~\beta}$, implying in turn the vanishing of $\dot{\mathbb R}^{\alpha \beta}$. 

Consider a congruence with zero shear, and set ${\mathbb R}^a_{\phantom{a}b}= K^a_{\phantom{a}b}$. Then, requiring $g^b=0$ imposes the following condition (refer Eq. (\ref{eq:gmag}), assuming $\nabla_{\bm n} \bm n=0$)
    \begin{eqnarray}
    - c^2 {\mathscr E}_n + \frac{\partial a_n}{\partial n} + \frac{a_n^2}{c^2}
    - \left(K^2\right)_{nn}=0
    \label{eq:cond-EKsq}
    \end{eqnarray} 
where $\left(K^2\right)_{nn}=\left(K^2\right)^a_{\phantom{a}m} n_a n^m$. The possibility of a rigidly rotating congruence is then tied to the existence of direction(s) $\bm n$ for which the above equation has solutions. More importantly, one can directly connect the rotation of tetrads suited to this congruence in terms of the Riemann tensor. An essential difference from the case of rectilinear accelerated motion is this connection is essentially pointwise, and there is no differential equation to be satisfied. We will illustrate all these points with some examples below.
\\
\\
\textit{Minkowski spacetime}: One immediate point to be noted is the impossibility to have pure vorticity for inertial congruences in Minkowski spacetime. One needs acceleration $a^i\neq0$ to generate rotational motion in Minkowski spacetime. For uniform circular motion in the $\theta=\pi/2$ plane of Minkowski spacetime (in spherical polar coordinates 
$(t, r, \theta, \phi)$), with angular velocity $\Omega$: $a^r=-\Omega^2 r/(1 - \Omega^2 r^2)$, and 
$\left(K^2\right)^r_{\phantom{a}r} = - \Omega^2/(1-\Omega^2 r^2)^2$. It is therefore easily verified that the above equation holds for $\bm n={\widehat {\bm r}}$. An alternate way of stating this result is that ${\widehat {\bm r}}$ is the only direction along which the congruence can look ``rigid", with zero relative acceleration between its members.
\\
\\
\textit{Schwarzschild spacetime}: In Schwarzschild spacetime, the situation changes drastically, for two reasons: (i) the curvature is non-zero, and (ii) the geometry depends on $r$. Consider again inertial frames ($a^i=0$) which are rotating, that is, whose four-velocity is parallel to $(1,0,0,\Omega)$. one now finds that shear associated with the above congruence is automatically zero, and $K^a_{\phantom{a}b}$ is easily computed. However, now we have a non-zero ${\mathscr E}^a_{\phantom{a}b}$, leading to the possibility that rotating, inertial frames can be supported by curvature. To have zero acceleration, one requires $\Omega=\Omega_0=\sqrt{(M/r^3)}$. It is easy to find the eigenvectors of ${\mathscr E}^a_{\phantom{a}b}$ and $-(K^2)^a_{\phantom{a}b}$ and identify if any of their common eigen-vectors have the same eigenvalue, thereby satisfying the above condition. The form of these matrices, in standard Schwarzschild coordinates $(t,r,\theta,\phi)$ can be easily found. 
The only common eigenvector of these matrices, whose eigenvalues satisfy the condition in Eq. (\ref{eq:cond-EKsq}), is 
\footnote{An interpretation of this vector is easily obtained in terms of a Fermi-Walker transported frame $\left( {\bm e}_{(0)}, {\bm e}_{(1)}, {\bm e}_{(2)}, {\bm e}_{(3)} \right)$ given in \cite{parker-pimentel}. In terms of the Fermi frame, which is used as the local \textit{compass of inertia}, we have ${\widehat{\bm r}}={\bm e}_{(1)} \cos{\Omega_0 \tau} + {\bm e}_{(3)} \sin{\Omega_0 \tau}$ and ${\bm n}_\star=-{\bm e}_{(1)} \sin{\Omega_0 \tau} + {\bm e}_{(3)} \cos{\Omega_0 \tau}$. Hence, $({ {\bm r}}, {\bm n}_\star)$ rotate with an angular velocity $\Omega_0$ with respect to the Fermi frame.
}
\begin{eqnarray}
    n_\star^i=\left(\frac{r \Omega_0}{q(r)},0,0,\frac{q(r)}{r-3 M}\right)
\end{eqnarray}
where $q(r)=\sqrt{(1-2M/r)(1-3M/r)}$. The corresponding eigenvalue is ${\mathscr E}_{n}=- \left(K^2\right)_{nn}=M/r^3$; that is $\Omega_0=\sqrt{{\mathscr E}_n}$. 
\\\\
The novelty of our derivation lies in the fact that we have related $\Omega_0$ as well as the direction in which ``rigidity" can be maintained, directly in terms of the Riemann tensor. Note the stark contrast with rotation in Minkowski spacetime, which necessarily requires acceleration to support it. The situation here is exactly reversed: $\widehat{\bm r}$ direction is no longer rigid, while the ``angular" direction $\bm n$ is.
%
%
\\\\
\noindent \textit{\textbf{Outlook and Discussion:}} The most common understanding of the equivalence principle is that one can ``mimic" gravity by going to a ``locally" accelerated frame of reference in Minkowski. This statement becomes extremely confusing when one realizes that gravity is encoded in the Riemann tensor, while the usual accelerated frame one chooses is the Rindler frame in Minkowski spacetime. This frame is rigid in flat spacetime, but not in a curved spacetime. If one uses rigid frames in curved spacetimes, the corresponding acceleration profile in Minkowski spacetime differ from Rindler through terms determined by electric part of Riemann. Individual timelike curves of the congruence can still be uniformly accelerated, but the change in acceleration from one curve to the next encodes information about the Riemann tensor. Explicitly deriving an expression for this, using the deviation equation between members of a general congruence, is one key result of this work. This has implications for conventional applications of the principle of equivalence, as well as for the Mach principle; the results and ideas presented in this work points to an intimate connection between inertial and non-inertial frames in curved spacetime, equivalence principle, and Mach principle
\footnote{A discussion of inertial frames along these lines can be found in \cite{pirani-inertial}}.
\\ \\
In particular, the results might also help characterize better the historically notorious notion of a frame that can represent a \textit{uniform gravitational field}. A natural characterization of such a frame would be in terms of a non-rotating congruence for which $a_n'\equiv \partial a_{n}/\partial n=0$, $a_n=a_0$ and that would imply (see Eq. (\ref{eq:gmag}))
\begin{eqnarray}
       \frac{1}{Z} \frac{d^2 Z}{d \tau^2} =
    - c^2 {\mathscr E}_n + \frac{a_0^2}{c^2} 
    \label{eq:nkd-UGF}
\end{eqnarray}
where we have assumed $\nabla_{\bm n} \bm n=0$. One can not have a rigid frame (LHS=0) unless ${\mathscr E}_n=$ constant, while for ${\mathscr E}_n=$ constant $={\mathscr E}_0$, a rigid frame would correspond to 
\begin{eqnarray}
       a_0 = c^2 \sqrt{{\mathscr E}_0}
\end{eqnarray}
This has no real solutions for ${\mathscr E}_0<0$. Some particularly interesting examples were considered in a similar spirit in \cite{naresh} for spherical symmetry, with comments on implications for Mach principle.
\\\\
Finally, moving over to the quantum domain, our result acquires importance in trying to understand the role of curvature in vacuum processes such as Unruh effect and Schwinger effect. The key reason for this is the well known, fundamental role played by \textit{shearing of trajectories}, that can be described geometrically using the deviation equation, in these processes. A direct demonstration of this was recently given in \cite{hari-dk-acc-detectors}, where it was shown that transition rate of an accelerated Unruh-de Witt detector, coupled to a quantum field in a Hadamard state in arbitrary curved spacetime, is Planckian to leading order with temperature 
$$
k_B T_{\rm Unruh} = \frac{\hbar \sqrt{a^2 - c^4 \mathscr{E}_n}}{2 \pi c}
$$
if effects due to off the plane components and derivatives of Riemann are ignored. Our result in this paper gives a novel interpretation to the above (rigorously obtained) result. If the accelerated trajectory $C_0$ in question is a member of a rigid congruence, we can write this using our result as 
$$
k_B T_{\rm Unruh} = \frac{\hbar \sqrt{-[a_n']_{C_0}}}{2 \pi}
$$ 
where $[a_n']_{C_0}$ denotes $\partial a_n/\partial n$ evaluated on the member $C_0$ of the congruence
\footnote{
The above result is exact in (anti-) de Sitter spacetime, for which $\mathscr{E}_n=-\Lambda$ and we obtain
$$
k_B T_{\rm Unruh} = \frac{\hbar \sqrt{a^2 + c^4 \Lambda}}{2 \pi c}
$$ 
}.
Despite the difficulties involved in identifying temperature in an arbitrary curved spacetime for an arbitrary accelerated detector, the above result is highly suggestive. It might be the first step in discerning the role of {\it relative}, as opposed to {\it absolute}, acceleration in physical effects in curved spacetime. As such, it might have a deeper significance for semi-classical gravity, and is worth investigating further.
\\\\
Finally, the stand-alone significance of the generalised, scalar version of the geodesic deviation equation of GR (with acceleration and rotation incorporated) must also be highlighted. Deviation equation is of central importance in conceptual as well as observational research in gravitational physics, and lies at the heart of operational significance of Riemann tensor. We have here extended it in a manner that also establishes its crucial role in determining non-inertial frames in curved spacetimes. Even most modern results such as gravitational wave memory, black hole shadows and particle production in curved spacetimes are rooted in deviation equation. Specifically, quantum corrections to our equation should have observable effects that can be detected with upcoming gravitational wave detectors. It should be also be useful in investigating the fate of spacetime singularities, along the lines of \cite{dk-bgv}.
\\\\
\textit{Acknowledgments:} I am grateful to Ghanshyam Date, Sayan Kar, and Naresh Dadhich for providing insightful and critical comments on the manuscript, and to Naresh for pointing out Ref. \cite{naresh}. I would also like to acknowledge important inputs and suggestions from Hari K. and Mayank Vashishtha. All numerical computations in this work have been done in \texttt{MATHEMATICA}.

\newpage

\appendix
\section{Supplemental Material - Deviation equation} 
\label{app:deviation-deriv}

Let us first state a few results for Fermi derivative along a given curve, defined by
\begin{eqnarray}
    \frac{D_F q^i}{d \tau} = \nabla_{\bm u} q^i + (a^i u_j - u^i a_j) q^j
\end{eqnarray}
Clearly, 
\begin{eqnarray}
    \frac{\widetilde D_F q^i}{d \tau} = \frac{D_F q^i}{d \tau} - {\mathbb R}^i_{\phantom{i} j} q^j
\end{eqnarray}
One can now prove the following identity for any vector $\bm x$ which is orthogonal to $\bm u$; that is $\bm x \bm \cdot \bm u=0$:
\begin{eqnarray}
    \frac{D_F^2 \bm x}{d \tau^2} = \left( \nabla^2_{\bm u} \bm x \right)_{\perp} - (\bm x \cdot \bm a) \bm a
\end{eqnarray}
The above relation can be obtained by using the definition of Fermi derivative twice and using simplifications that follow from repeated differentiation of the condition $\bm x \bm \cdot \bm u=0$ along $\bm u$. 

Another useful identity one may prove is the following: For any vector $\bm y$ such that $\mathcal{L}_{\bm u} \bm y=\bm b$, we have 
\begin{eqnarray}
    \nabla^2_{\bm u} \bm y = u^k y^l \left[\nabla_k, \nabla_l \right] \bm u + \nabla_{\bm y} \bm a + \left(\nabla_{\bm u} \bm b + \nabla_{\bm b} \bm y \right)
\end{eqnarray}
It is easy to see that the above identities yield the deviation equation for a non-rotating tetrad as given in Hawking and Ellis \cite{hawking-ellis}.

Now, given the vectors $Z^a$ and $\xi^a$ as defined in the main text, we have 
\begin{eqnarray}
    \nabla_{\bm u} \bm Z &=& \nabla_{\bm u} \bm \xi + (\bm \xi \cdot \bm u) \bm a + (\bm \xi \cdot \bm a) \bm u
    \nonumber \\
    &=& \nabla_{\bm \xi} \bm u + (\bm \xi \cdot \bm u) \bm a + (\bm \xi \cdot \bm a) \bm u
    \nonumber \\
    &=& Z^k \underbrace{\left(\nabla_k u^m + u_k a^m \right)}_{K^m_{\phantom{m}k}} {\bm e_{(m)}} + (\bm Z \cdot \bm a) \bm u
\end{eqnarray}
where the last step follows by noting that $K^m_{\phantom{m}k} u^k=0$.

The generalised equation, with rotation included, can be derived essentially by using the above identities, although the derivation is somewhat lengthier and we skip the details. We only state one particular identity involving the rotation matrix which is required to arrive at the final expression (see text for definitions of symbols): 
\begin{eqnarray}
    Z_m \overset{\perp}{\nabla}_{\bm u} {\mathbb R}^{am} = e^a_{\alpha} e^m_{\beta} \dot{\mathbb R}^{\alpha \beta} 
\end{eqnarray}
\\ where the $\perp$ refers to orthogolaization on the index $a$.
To prove the above, we expand $R^{am}$ in the tetrad basis, and use the conditions ${\widetilde D_F e^a_{\alpha}}/{d \tau}=0$ to simplify further. All the terms quadratic in rotation matrices cancel out, yielding the RHS above.

\section{The form of the matrices in Sec V, Schwarzschild} 
Explicit form of the matrices, in standard Schwarzschild coordinates $(t,r,\theta,\phi)$, whose eigen-vectors and eigen-values are quoted in Sec V, Schwarschild spacetime.
\begin{widetext}
\begin{eqnarray}
{\mathscr E}^a_{\phantom{a}b} &=& \frac{1}{r-3M} \left(
\begin{array}{cccc}
 -M \Omega_0^2 & 0 & 0 & M \Omega_0 \\
 0 & - \Omega_0^2 (2 r-3 M) & 0 & 0 \\
 0 & 0 & \frac{M}{r^2} & 0 \\
 (2 M-r) \Omega_0^3 & 0 & 0 & -\Omega_0^2 (2 M-r) \\
\end{array}
\right)  
\nonumber \\
\nonumber \\
\nonumber \\
(K^2)^a_{\phantom{a}b} &=& \frac{1}{r-3M}
\left(
\begin{array}{cccc}
 M \Omega_0^2 & 0 & 0 & -\Omega_0^3 \\
 0 & -\Omega_0^2(r-3M) & 0 & 0 \\
 0 & 0 & 0 & 0 \\
 -(2 M-r) \Omega_0^3 & 0 & 0 & \Omega_0^2 (2M-r) \\
\end{array}
\right)
\end{eqnarray}     
\end{widetext}

\bibliography{apssamp}

\providecommand{\noopsort}[1]{}\providecommand{\singleletter}[1]{#1}%
\begin{thebibliography}{14}%
\makeatletter
\providecommand \@ifxundefined [1]{%
 \@ifx{#1\undefined}
}%
\providecommand \@ifnum [1]{%
 \ifnum #1\expandafter \@firstoftwo
 \else \expandafter \@secondoftwo
 \fi
}%
\providecommand \@ifx [1]{%
 \ifx #1\expandafter \@firstoftwo
 \else \expandafter \@secondoftwo
 \fi
}%
\providecommand \natexlab [1]{#1}%
\providecommand \enquote  [1]{``#1''}%
\providecommand \bibnamefont  [1]{#1}%
\providecommand \bibfnamefont [1]{#1}%
\providecommand \citenamefont [1]{#1}%
\providecommand \href@noop [0]{\@secondoftwo}%
\providecommand \href [0]{\begingroup \@sanitize@url \@href}%
\providecommand \@href[1]{\@@startlink{#1}\@@href}%
\providecommand \@@href[1]{\endgroup#1\@@endlink}%
\providecommand \@sanitize@url [0]{\catcode `\\12\catcode `\$12\catcode `\&12\catcode `\#12\catcode `\^12\catcode `\_12\catcode `\%12\relax}%
\providecommand \@@startlink[1]{}%
\providecommand \@@endlink[0]{}%
\providecommand \url  [0]{\begingroup\@sanitize@url \@url }%
\providecommand \@url [1]{\endgroup\@href {#1}{\urlprefix }}%
\providecommand \urlprefix  [0]{URL }%
\providecommand \Eprint [0]{\href }%
\providecommand \doibase [0]{https://doi.org/}%
\providecommand \selectlanguage [0]{\@gobble}%
\providecommand \bibinfo  [0]{\@secondoftwo}%
\providecommand \bibfield  [0]{\@secondoftwo}%
\providecommand \translation [1]{[#1]}%
\providecommand \BibitemOpen [0]{}%
\providecommand \bibitemStop [0]{}%
\providecommand \bibitemNoStop [0]{.\EOS\space}%
\providecommand \EOS [0]{\spacefactor3000\relax}%
\providecommand \BibitemShut  [1]{\csname bibitem#1\endcsname}%
\let\auto@bib@innerbib\@empty
\bibitem [{\citenamefont {Synge}(1971)}]{synge-gr}%
  \BibitemOpen
  \bibfield  {author} {\bibinfo {author} {\bibfnamefont {J.}~\bibnamefont {Synge}},\ }\href {https://books.google.co.in/books?id=N6lKMQAACAAJ} {\emph {\bibinfo {title} {Relativity: The General Theory}}},\ North-Holland series in physics\ (\bibinfo  {publisher} {North-Holland Publishing Company},\ \bibinfo {year} {1971})\BibitemShut {NoStop}%
\bibitem [{\citenamefont {Misner}\ \emph {et~al.}(1973)\citenamefont {Misner}, \citenamefont {Thorne},\ and\ \citenamefont {Wheeler}}]{mtw}%
  \BibitemOpen
  \bibfield  {author} {\bibinfo {author} {\bibfnamefont {C.}~\bibnamefont {Misner}}, \bibinfo {author} {\bibfnamefont {K.}~\bibnamefont {Thorne}},\ and\ \bibinfo {author} {\bibfnamefont {J.}~\bibnamefont {Wheeler}},\ }\href@noop {} {\emph {\bibinfo {title} {{Gravitation}}}}\ (\bibinfo  {publisher} {W. Freeman},\ \bibinfo {year} {1973})\BibitemShut {NoStop}%
\bibitem [{\citenamefont {Hawking}\ and\ \citenamefont {Ellis}(1975)}]{hawking-ellis}%
  \BibitemOpen
  \bibfield  {author} {\bibinfo {author} {\bibfnamefont {S.~W.}\ \bibnamefont {Hawking}}\ and\ \bibinfo {author} {\bibfnamefont {G.~F.~R.}\ \bibnamefont {Ellis}},\ }\href {http://www.amazon.com/Structure-Space-Time-Cambridge-Monographs-Mathematical/dp/0521099064} {\emph {\bibinfo {title} {The Large Scale Structure of Space-Time (Cambridge Monographs on Mathematical Physics)}}}\ (\bibinfo  {publisher} {Cambridge University Press},\ \bibinfo {year} {1975})\BibitemShut {NoStop}%
\bibitem [{\citenamefont {Kothawala}(2020)}]{dk-bgv}%
  \BibitemOpen
  \bibfield  {author} {\bibinfo {author} {\bibfnamefont {D.}~\bibnamefont {Kothawala}},\ }\bibfield  {title} {\bibinfo {title} {{BGV} theorem, geodesic deviation, and quantum fluctuations},\ }\href {https://doi.org/10.1088/1361-6382/abd146} {\bibfield  {journal} {\bibinfo  {journal} {Classical and Quantum Gravity}\ }\textbf {\bibinfo {volume} {38}},\ \bibinfo {pages} {045006} (\bibinfo {year} {2020})}\BibitemShut {NoStop}%
\bibitem [{Note1()}]{Note1}%
  \BibitemOpen
  \bibinfo {note} {The consequences of such a condition for rigidity have been discussed in the literature previously (see, for example, \cite {synge-gr} and \cite {pirani-rigidity}), although these discussions mostly work only with $\protect \dot Z$ and not its second derivative.}\BibitemShut {Stop}%
\bibitem [{\citenamefont {Hari}\ and\ \citenamefont {Kothawala}(2021)}]{hari-dk-acc-detectors}%
  \BibitemOpen
  \bibfield  {author} {\bibinfo {author} {\bibfnamefont {K.}~\bibnamefont {Hari}}\ and\ \bibinfo {author} {\bibfnamefont {D.}~\bibnamefont {Kothawala}},\ }\bibfield  {title} {\bibinfo {title} {{Effect of tidal curvature on dynamics of accelerated probes}},\ }\href {https://doi.org/10.1103/PhysRevD.104.064032} {\bibfield  {journal} {\bibinfo  {journal} {Phys. Rev. D}\ }\textbf {\bibinfo {volume} {104}},\ \bibinfo {pages} {064032} (\bibinfo {year} {2021})},\ \Eprint {https://arxiv.org/abs/2106.14496} {arXiv:2106.14496 [gr-qc]} \BibitemShut {NoStop}%
\bibitem [{\citenamefont {Poisson}(2004)}]{poisson-toolkit}%
  \BibitemOpen
  \bibfield  {author} {\bibinfo {author} {\bibfnamefont {E.}~\bibnamefont {Poisson}},\ }\href@noop {} {\emph {\bibinfo {title} {A Relativist’s Toolkit: The Mathematics of Black-Hole Mechanics}}}\ (\bibinfo  {publisher} {Cambridge University Press},\ \bibinfo {year} {2004})\ \bibinfo {note} {ch. 2}\BibitemShut {NoStop}%
\bibitem [{Note2()}]{Note2}%
  \BibitemOpen
  \bibinfo {note} {An interpretation of this vector is easily obtained in terms of a Fermi-Walker transported frame $\left ( {\protect \bm {e}}_{(0)}, {\protect \bm {e}}_{(1)}, {\protect \bm {e}}_{(2)}, {\protect \bm {e}}_{(3)} \right )$ given in \cite {parker-pimentel}. In terms of the Fermi frame, which is used as the local \protect \textit {compass of inertia}, we have ${\setbox \z@ \hbox {\mathsurround \z@ $\textstyle \protect \bm {r}$}\mathaccent "0362{\protect \bm {r}}}={\protect \bm {e}}_{(1)} \cos {\Omega _0 \tau } + {\protect \bm {e}}_{(3)} \sin {\Omega _0 \tau }$ and ${\protect \bm {n}}_\star =-{\protect \bm {e}}_{(1)} \sin {\Omega _0 \tau } + {\protect \bm {e}}_{(3)} \cos {\Omega _0 \tau }$. Hence, $({ {\protect \bm {r}}}, {\protect \bm {n}}_\star )$ rotate with an angular velocity $\Omega _0$ with respect to the Fermi frame.}\BibitemShut {Stop}%
\bibitem [{Note3()}]{Note3}%
  \BibitemOpen
  \bibinfo {note} {A discussion of inertial frames along these lines can be found in \cite {pirani-inertial}}\BibitemShut {NoStop}%
\bibitem [{\citenamefont {Dadhich}(1997)}]{naresh}%
  \BibitemOpen
  \bibfield  {author} {\bibinfo {author} {\bibfnamefont {N.}~\bibnamefont {Dadhich}},\ }\bibfield  {title} {\bibinfo {title} {{On ``minimally curved spacetimes'' in general relativity}},\ }\href@noop {} {\bibfield  {journal} {\bibinfo  {journal} {preprint}\ } (\bibinfo {year} {1997})},\ \Eprint {https://arxiv.org/abs/gr-qc/9705026} {arXiv:gr-qc/9705026 [gr-qc]} \BibitemShut {NoStop}%
\bibitem [{Note4()}]{Note4}%
  \BibitemOpen
  \bibinfo {note} {The above result is exact in (anti-) de Sitter spacetime, for which $\protect \mathscr {E}_n=-\Lambda $ and we obtain $$ k_B T_{\protect \rm Unruh} = \protect \frac {\hbar \protect \sqrt {a^2 + c^4 \Lambda }}{2 \pi c} $$}\BibitemShut {NoStop}%
\bibitem [{\citenamefont {Pirani}\ and\ \citenamefont {Williams}(1962)}]{pirani-rigidity}%
  \BibitemOpen
  \bibfield  {author} {\bibinfo {author} {\bibfnamefont {F.~A.~E.}\ \bibnamefont {Pirani}}\ and\ \bibinfo {author} {\bibfnamefont {G.}~\bibnamefont {Williams}},\ }\bibfield  {title} {\bibinfo {title} {Rigid motion in a gravitational field},\ }\href {http://www.numdam.org/item/SJ_1961-1962__5__A8_0/} {\bibfield  {journal} {\bibinfo  {journal} {Seminaire Janet. Mecanique analytique et mecanique celeste}\ }\textbf {\bibinfo {volume} {5}},\ \bibinfo {pages} {1} (\bibinfo {year} {1961-1962})}\BibitemShut {NoStop}%
\bibitem [{\citenamefont {Parker}\ and\ \citenamefont {Pimentel}(1982)}]{parker-pimentel}%
  \BibitemOpen
  \bibfield  {author} {\bibinfo {author} {\bibfnamefont {L.}~\bibnamefont {Parker}}\ and\ \bibinfo {author} {\bibfnamefont {L.~O.}\ \bibnamefont {Pimentel}},\ }\bibfield  {title} {\bibinfo {title} {Gravitational perturbation of the hydrogen spectrum},\ }\href {https://doi.org/10.1103/PhysRevD.25.3180} {\bibfield  {journal} {\bibinfo  {journal} {Phys. Rev. D}\ }\textbf {\bibinfo {volume} {25}},\ \bibinfo {pages} {3180} (\bibinfo {year} {1982})}\BibitemShut {NoStop}%
\bibitem [{\citenamefont {Pirani}(1956)}]{pirani-inertial}%
  \BibitemOpen
  \bibfield  {author} {\bibinfo {author} {\bibfnamefont {F.~A.~E.}\ \bibnamefont {Pirani}},\ }\bibfield  {title} {\bibinfo {title} {On the definition of inertial systems in general relativity},\ }\href@noop {} {\bibfield  {journal} {\bibinfo  {journal} {Mercier and Kervaire}\ ,\ \bibinfo {pages} {198}} (\bibinfo {year} {1956})}\BibitemShut {NoStop}%
\end{thebibliography}%

\end{document}